\documentclass[usenatbib]{mn2e}
\usepackage{color}
\usepackage{ulem}

\usepackage{epsf}
\usepackage{graphicx}
\def\gsim { \lower .75ex \hbox{$\sim$} \llap{\raise .27ex \hbox{$>$}} }
\def\lsim { \lower .75ex \hbox{$\sim$} \llap{\raise .27ex \hbox{$<$}} }

\begin{document}

\title[Baryon effects on DM halos]{Baryon effects on the dark matter halos 
constrained from strong gravitational lensing} 
\author[Wang et al.]{
\parbox{\textwidth}{Lin Wang$^{1,2}$\thanks{E-mail: wl010@bao.ac.cn}, 
Da-Ming Chen$^{1,2}$, Ran Li$^1$}
\vspace*{4pt} \\
$^1$National Astronomical Observatories, Chinese Academy of Sciences, 20A Datun 
Road, Chaoyang District, Beijing 100012, China; \\
$^2$School of Astronomy and Space Science, University of Chinese Academy of 
Sciences, Beijing 100049, China }
\maketitle

\begin{abstract}
Simulations are expected to be the powerful tool to 
investigate the baryon effects on dark matter (DM) halos. Recent high  
resolution, cosmological hydrodynamic simulations (\citealt{Cintio14}, DC14) 
predict that the inner density profiles of DM halos depend systematically on 
the ratio of stellar to DM mass ($M_{\ast}/M_{\rm halo}$) which is thought to 
be able to provide good fits to the observed rotation curves of galaxies. 
The DC14 profile is fitted from the simulations which are confined 
to $M_{\rm halo}\le 10^{12}M_{\sun}$, in order to investigate the physical 
processes that may affect all halos, we extrapolate it to much larger 
halo mass, including that of galaxy clusters. The inner slope of DC14 profile 
is flat for low halo mass, it approaches 1 when the halo mass increases 
towards $10^{12}M_{\sun}$ and decreases rapidly after that mass. We use DC14 
profile for lenses and find
that it predicts too few lenses compared with the most recent strong lensing 
observations SQLS (\citealt{Inada12}). We also 
calculate the strong lensing probabilities for a simulated density profile which 
continues the halo mass from the mass end of DC14 ($\sim 10^{12}M_{\sun}$) to 
the mass that covers the galaxy clusters (\citealt{Schaller15}, Schaller15), and 
find that this Schaller15 model predict too many lenses compared with other 
models and SQLS observations. Interestingly, Schaller15 profile 
has no core, however, like DC14, the rotation curves of the simulated halos are 
in excellent agreement with observational data. Furthermore, we show that the 
standard two-population model SIS+NFW cannot match the most recent SQLS 
observations for large image separations. 

\end{abstract}
\begin{keywords}
Cosmology: theory---dark matter---galaxies: halos---gravitational 
lensing: strong
\vspace*{-0.5 truecm}
\end{keywords}





%


\section{Introduction}
\label{sec:intro}
It is well-known that General Relativity (GR) is very successful on small 
scales like our solar system.  When applied to scales of galaxies and larger, 
however, 
some exotic ingredients are needed to explain our observations. For 
example, when GR (its weak gravitational field form) is applied to 
galaxies, we need cold dark matter (CDM) to explain the 
observational data of rotation curves; when applied to cosmology, 
we need dark energy (DE) to explain the on going accelerating expansion of our 
Universe. To account for  the cosmic structure formation and gravitational 
lensing in the context of $\Lambda$CDM cosmology, both CDM and DE are needed.  
There are no direct observational evidences supporting the existence of CDM and 
DE. Their properties are assumed so that the usual  astronomical 
observations can be interpreted reasonably based on GR (and thus Newtonian 
theory of gravity).

In this paper, we focus on the properties of CDM, and consider only the 
observational constraints arising from rotation curves and strong lensing. 
Early optical and 21 cm line 
of neutral hydrogen observations for late-type disk galaxies all indicate the 
property of having an almost constant rotation velocity in their outer parts.
If Newtonian theory of gravity is correct, the flat rotation curves suggest the 
existence of some non-baryonic matter, called dark matter (DM), surrounding 
each observable galaxy as a dark halo. Other 
observations and structure formation theory require that DM is cold, that is, 
the DM particles are massive and their random velocity  is small. Furthermore, 
it turns out that the amount of CDM is at least several times larger in mass 
than observable baryonic matter. Therefore, the total density 
profiles of 
galaxies and clusters of galaxies are CDM dominated, the usual baryoic matter 
(usually resides in the central region) can play the role of changing the inner
slope, the importance of which depends on the amount it contributes to the 
total mass \citep{Schaller15a, Xu16, Yannick17}. For later 
considerations, we 
use the density profiles of DM halos 
to stand for the total mass distributions. The observational data for flat 
rotation curves can be 
well-described if the mass density profile of CDM particles is modeled as the
singular isothermal sphere (SIS): $\rho\sim r^{-2}$, when we 
observe the outer parts of disk galaxies. Interestingly, this steep, 
singular power-law model is preferred by strong gravitational lensing for giant 
elliptical galaxies.
On the other hand, however, recent high-resolution rotation velocity 
associated with dark matter in the inner parts of disk galaxies indicates the 
presence of constant density DM cores. In fact, it is now established that the 
cored isothermal sphere (CIS) fit well the observed rotation curves, both 
in the inner and outer parts of disk galaxies \citep{deBlok10},
\begin{equation}
 \rho_{CIS}(r)=\frac{\rho_0}{1+(r/r_c)^2},
\end{equation}
where $\rho_0$ is the central density, and $r_c$ is the core radius of a halo.
Unfortunately, CIS model cannot match strong lensing observations. 

Gravitational lensing provides a powerful tool to detect DM, although 
it is not sensitive to whether the mass doing the lensing is baryonic or 
dark, 
but rather simply depends on the total. For a certain given mass of a lensing 
galaxy or galaxy cluster, strong lensing efficiency is very sensitive to the 
slope $\gamma$ of the central total mass density profile ($\rho\propto 
r^{-\gamma}$).  For example, a cored density profile like CIS for a reasonable 
value of the core radius $r_c$ (usually determined by rotation curves) would 
lead to an extremely low lensing rate compared with lensing observations, 
while singular isothermal sphere 
(SIS, $\gamma=2$, for elliptical 
galaxies) matches observations well. As for galaxy clusters as lenses, NFW 
(\citealt{nfw96,nfw97};
$\gamma=1$) has been used as a good model, but as will be demonstrated in this 
paper, the most recent strong lensing observations require a steeper slope.
It should be pointed out that dark matter halos are triaxial rather 
than spherical \citep{Jing02, Despali14, Bonamigo15, Despali17}, the 
ellipticity would significantly increase the lensing efficiency 
\citep{Bartelmann98, Meneghetti01, Meneghetti03, Hennawi07, Broadhurst08}, but 
not so important compared with the inner slope. For example, \cite{Giocoli12} 
present MOKA, a new algorithm for simulating the gravitational lensing signal 
from cluster-sized halos, and find that the strong lensing cross sections 
depend most strongly on the concentration and on the inner slope of the density 
profile of a halo, followed in order of importance by halo triaxiality and the 
presence of a bright central galaxy.

The observations of rotation curves and strong lensing can only be used to 
constrain the density profile of halos, not directly the properties of CDM 
particles. It is the structure formation theory, mainly through computer 
simulations, that determines what the properties of CDM should be assumed so 
that the density profile can be correctly predicted. For example, if CDM 
is self-interacting, or DM particles are warm (e.g., 
\citealt{SGTF12}), a cored density profile can be created even
without baryons. In the standard, 
hierarchical, CDM paradigm of cosmological structure formation, galaxy 
formation begins with the gravitational collapse of
over dense regions into bound, virialized halos of CDM.  In this $\Lambda$CDM 
paradigm, halos form from purely 
collisionless DM particles with primordial power spectrum of fluctuations 
predicted by inflationary model. Small halos are the first to form, and larger 
halos form subsequently by mergers of pre-existing halos and by accretion 
of diffuse dark matter that has never been part of a halo. In a simplified 
picture (\citealt{WR78}), baryonic gas is initially well mixed with the DM 
particles, then participates in the gravitational collapse of DM and is heated 
by shocks to the virial temperature of the DM halos.  Bound in
the potential wells of DM halos, baryonic gas proceed to cool radiatively due 
to bremsstrahlung, recombination and collisionally exited line emission 
\citep{FW12}. 

A full analytic description of the development of such dissipationless 
hierarchical clustering came in the ealy 1990's with extensions of the original 
Press-Schechter model based on excursion set theory 
(\citealt{Bower91,BCEK91,LC93,KW93}). The halo mass function derived in this 
analytic theory is in well aggreement with that from DM only simulations 
\citep{Sheth99, Jenkins01, Warren06, Reed07, Tinker08, Crocce10, 
Courtin10, Angulo12, Watson13, Despali16}. We need such mass function in our 
lensing probability calculations. 

The current computer simulations are the most robust tools to explore the 
formation and evolution of the large scale structure of the universe 
\citep{FW12}. In the 
$\Lambda$CDM paradigm, purely CDM N-body simulations can reproduce the 
observed cosmic web as demonstrated by Sloan Digital Sky Survey (SDSS). In this 
scenario, observable galaxies made-up of baryons form at the centers of DM 
halos. Unfortunately, the cosmic gas (the initial form of baryons), and the 
subsequent star formation processes, are poorly understood. Based on N-body 
technique, there are mainly two complementary methods for simulating the galaxy 
formations. The direct inclusion of the baryonic component and all the 
astrophysical processes affecting it, known as numerical hydrodynamic method, 
is too computationally taxing to perform for large samples of galaxies 
\citep{Teyssier02, Springel10a, Springel10b, Commercon14, 
Tescari14, Pakmor16, Katsianis17}. This 
method can treat reliably only a subset of the relevant gas physics, such as 
the shock heating of gas and its subsequent radiative cooling. Another method 
is known as semi-analytic modeling. The main difference with the direct 
simulation is that, instead of solving the equations of hydrodynamics directly, 
one employs a simple, spherically symmetric model in which the gas is assumed 
to have been fully shock-heated to the virial temperature of each halo, so that 
its cooling and accretion can be accurately calculated. This phenomenological 
treatments of baryonic processes are based on physical insights gained from 
simulations of individual systems and from observations. Uncertainty parameters 
such as the efficiency of star formation and stellar feedback can be adjusted 
to reproduce the observed properties of all types of galaxies and clusters of 
galaxies. 

Over the past decades, a range of studies have associated
galaxies with DM halos at a given epoch, using a variety of techniques,
including halo occupation distribution medeling (e.g., \citealt{BW02,
Bullock02}), the conditional luminosity function modeling (e.g.,
\citealt{Yang03}), and variants of the abundance matching technique (e.g.,
\citealt{KK99, Ney04, Tas04, Con06}), which relate the 
observed properties of galaxies to their formation histories in a hierarchical 
manner. 

Recent high resolution, cosmological hydrodynamic simulations 
(\citealt{Cintio14}, DC14) introduce a mass-dependent density 
profile to describe the distribution of dark matter within galaxies, which 
takes into account the stellar-to-halo mass ratio ($M_\ast/M_{\rm halo}$) 
dependence of baryon effects on DM. Using a Markov Chain Monte 
Carlo approach, \citet{Katz16} found that the DC14 model provides better fits 
to the most recent observed rotation curves of galaxies over a large range of 
luminosity and surface brightness than do halo models which neglect baryonic 
physics (i.e., NFW). 
 
In this paper we  apply the DC14 model to 
the strong lensing probability calculations and compare the results with 
the most recent strong lensing observations. Although the most 
recent rotation curves for disk galaxies support the D14 model, however, there 
are no direct evidences arising from the simulations show us that all the 
central galaxies of the final halos are disk galaxies.  In fact, similar to the 
previous work in the literature (e.g., \citealt{SIR99}, \citealt{MCW06}, 
\citealt{MWC08}), the morphological types of galaxies are unconcerned in these 
simulations. The combined baryon effects such as gas cooling, stellar feedback 
and dynamical friction are modeled to reshape the inner slope of the density 
profile of DM halos which initially have the functional form of NFW, rather 
than to distinguish disk galaxies from ellipticals. The main 
difference with previous similar work is that DC14 profile depends on the halo 
mass. At low mass end, each halo display a central core, and for
halos with increasing mass,  some astrophysical processes erase the central 
cores and steepen the inner slopes of the DM density profiles. So it would be 
interesting to check whether or not the  DC14 model, which is centrally 
steepened for $M_{\rm halo}\sim 10^{12}M_{\sun}$, is able to 
describe the massive galaxies. 

Clearly, it would be very helpful for us to understand the baryon effects on DM 
distributions if we have a simulated density profile which continues the halo 
mass from the mass end of DC14 ($\sim 10^{12}M_{\sun}$) to the mass that covers
the galaxy clusters.   One such 
example is the investigation for the internal structure and density profiles of 
halos of mass $10^{10}-10^{14}M_{\sun}$ in the Evolution and Assembly of 
Galaxies and their Environment (EAGLE) simulations (\citealt{Schaller15}, 
Schaller15). These 
follow the formation of galaxies in a $\Lambda$CDM cosmology and include a 
treatment of the baryon physics thought to be relevant. As desired, in this 
mass range the total density profile is similar to NFW in the inner and outer 
parts, but has a slope of $-2$ at some radius $r_{\rm i}\sim 2.27$kpc 
(approximately independent of the total mass) relatively near the centers of 
halos. We calculate the lensing probabilities corresponding to the Schaller15 
model and compare the results with other models and observations.

For comparisons, we also demonstrate the lensing 
probabilities for SIS + NFW and DC14($\beta$ = $\gamma$ =2) +NFW models. 
We compare these results with observation of JVAS/CLASS survey and 
the Sloan Digital Sky Survey Quasar Lens Search (SQLS, \citealt{Inada12}). 
We adopt the most generally accepted values of the parameters for flat 
$\Lambda$CDM cosmology, for which, with usual symbols,  $\Omega_{\mathrm 
m}=0.27$, $\Omega_{\Lambda}=0.73$, $h=0.75$ and $\sigma_8=0.8$.

This paper is organized as follows: in Section~\ref{sec:lensing equation} we 
present the lensing equations for DC14 and Schaller15 models. We calculate the 
lensing 
probabilities for different profiles and compare them 
with observations in Section~\ref{sec:probab}. The discussions and
conclusions are presented in Section~\ref{sec:conc}.

 \section{Lensing equations}
\label{sec:lensing equation}
\subsection{DC14 model}

 The DC14 model is derived by fitting the so-called $(\alpha,\beta,\gamma)$ 
double power-law model to the simulations, the resultant density profile is

\par
\begin{equation}
{\rho(r)}= \frac{\rho_{\rm s}}{\left(\frac{r}{r_{\rm 
s}}\right)^{\gamma}\left[1+\left(\frac{r}{r_{\rm 
s}}\right)^{\alpha}\right]^{(\beta-\gamma)/\alpha}},
\label{eq:DC14}
\end{equation}
where $\rho_\mathrm{s}$ is the scale density and $r_\mathrm{s}$ the scale 
radius, and
\begin{eqnarray}
\alpha&=& 2.94-\log_{10}[(10^{X+2.33})^{-1.08}+(10^{X+2.33})^{2.29}] 
\nonumber \\
\beta&=&4.23+1.34X+0.26X^2 \\
\gamma&=&-0.06+\log_{10}[(10^{X+2.56})^{-0.68}+(10^{X+2.56})] \nonumber
\end{eqnarray}
\label{abg}
where $X=\log_{10}(M_\ast/M_{\rm halo})$ and the mass range 
of validity of 
$\alpha$,$\beta$, and $\gamma$ is $-4.1 <\log_{10}(M_\ast/M_{\rm 
halo})<-1.3$. For our purpose, we need to know 
$\alpha$, $\beta$ and $\gamma$ as functions of the halo mass $M_{\rm halo}$. 
Fortunately,  we have a good fitting formula at hand (\citealt{guo10} )
\begin{equation}
M_{\ast}/M_{\rm halo}=0.129\left[\left(\frac{M_{\rm halo}}{M_0}\right)^{-0.926}
+\left(\frac{M_{\rm halo}}{M_0}\right)^{0.261}\right]^{-2.440}
\label{guo}
\end{equation}
where $M_{0}=10^{11.4}M_{\sun}$. This formula is valid when the halo mass 
ranges from $10^{10.8}M_{\sun}$ to $10^{14.9}M_{\sun}$, a range that dominates 
the strong lensing probabilities.

As usual, we define the mass of a halo to be the mass within $r_{200}$ (which 
is the radius of a sphere around a DM halo within which the average mass 
density is $200$ times the critical mean mass density of the universe),
\begin{equation}
M_\mathrm{halo}=4\pi\int^{r_{200}}_0\rho
r^2dr=4\pi\rho_\mathrm{s}r_\mathrm{s}^3f(c_1), \label{mdm}
\end{equation}
with $c_1=r_{200}/r_\mathrm{s}$ the concentration parameter, and
\begin{equation}
f(c_1)=\int^{c_1}_0\frac{x^2dx}{x^{\gamma}(1+x^{\alpha})^{(\beta-\gamma)/\alpha
}
}.
\end{equation}
In flat $\Lambda$CDM cosmology, the parameters $\rho_\mathrm{s}$ and 
$r_\mathrm{s}$ can  be expressed as (\citealt{li02,Chen03a}),
\begin{equation}
\rho_\mathrm{s}=\rho_\mathrm{crit}\left[\Omega_\mathrm{m}(1+z)^3
+\Omega_{\Lambda}\right]\frac{200}{3}\frac{c_1^3}{f(c_1)},
\end{equation}

\begin{equation}
r_\mathrm{s}=\frac{1.626}{c_1}\frac{M_{15}^{1/3}}
{\left[\Omega_\mathrm{m}
(1+z)^3+\Omega_{\Lambda}\right]^{1/3}}h^{-1}\mathrm{Mpc}.
\label{rs}
\end{equation}
where $\rho_{crit}$ is the present value of the critical mass density of the 
universe, and $M_{15}$ is the reduced mass of a halo defined as
$M_{15}=M_\mathrm{halo}/(10^{15}\mathrm{h}^{-1}M_{\sun})$.

The surface mass density for the DC14 profile is
\begin{eqnarray}
    \Sigma(x) = 2\rho_s r_s V(x)
    \label{sur_DC14}
\end{eqnarray}
where
\[
V(x)=\int_0^{\infty}\left(x^2+z^2\right)^{-\gamma/2}
       \left[\left(x^2+z^2\right)^{\alpha/2}+1\right]^{(\gamma-\beta)/\alpha}dz,
\]
and $x=|\vec{x}|$, $\vec{x} = \vec{\xi}/r_s$, $\vec{\xi}$ is the position
vector in the lens plane.  We thus obtain the lensing equation for a 
DC14 halo
\begin{eqnarray}
    y = x - \mu_s {g(x)\over x}\,,
    \label{lens1}
\end{eqnarray}
where $y = |\vec{y}|$, $\vec{\eta} = \vec{y}\, r_s D_\mathrm{S}/D_\mathrm{L}$ 
is the position 
vector in the source plane, and
\begin{eqnarray}
    g(x) \equiv \int_0^x u V(u) du,
    \label{gx}
\end{eqnarray}
and
\begin{eqnarray}
    \mu_s \equiv {4\rho_s r_s\over \Sigma_{\rm cr}}\,,
    \label{mus}
\end{eqnarray}
where $\Sigma_\mathrm{cr}=(c^2/4\pi
G)(D_\mathrm{S}/D_\mathrm{L}D_\mathrm{LS})$ is
the critical surface mass density; $D_\mathrm{L}$, $D_\mathrm{S}$  and
$D_\mathrm{LS}$ are the angular diameter distances from the observer to the 
lens, to the source and from the lens to the source, respectively.

\begin{figure}
\begin{center}
\mbox{
\hspace{-0.7cm}
\resizebox{6.7cm}{!}{\includegraphics{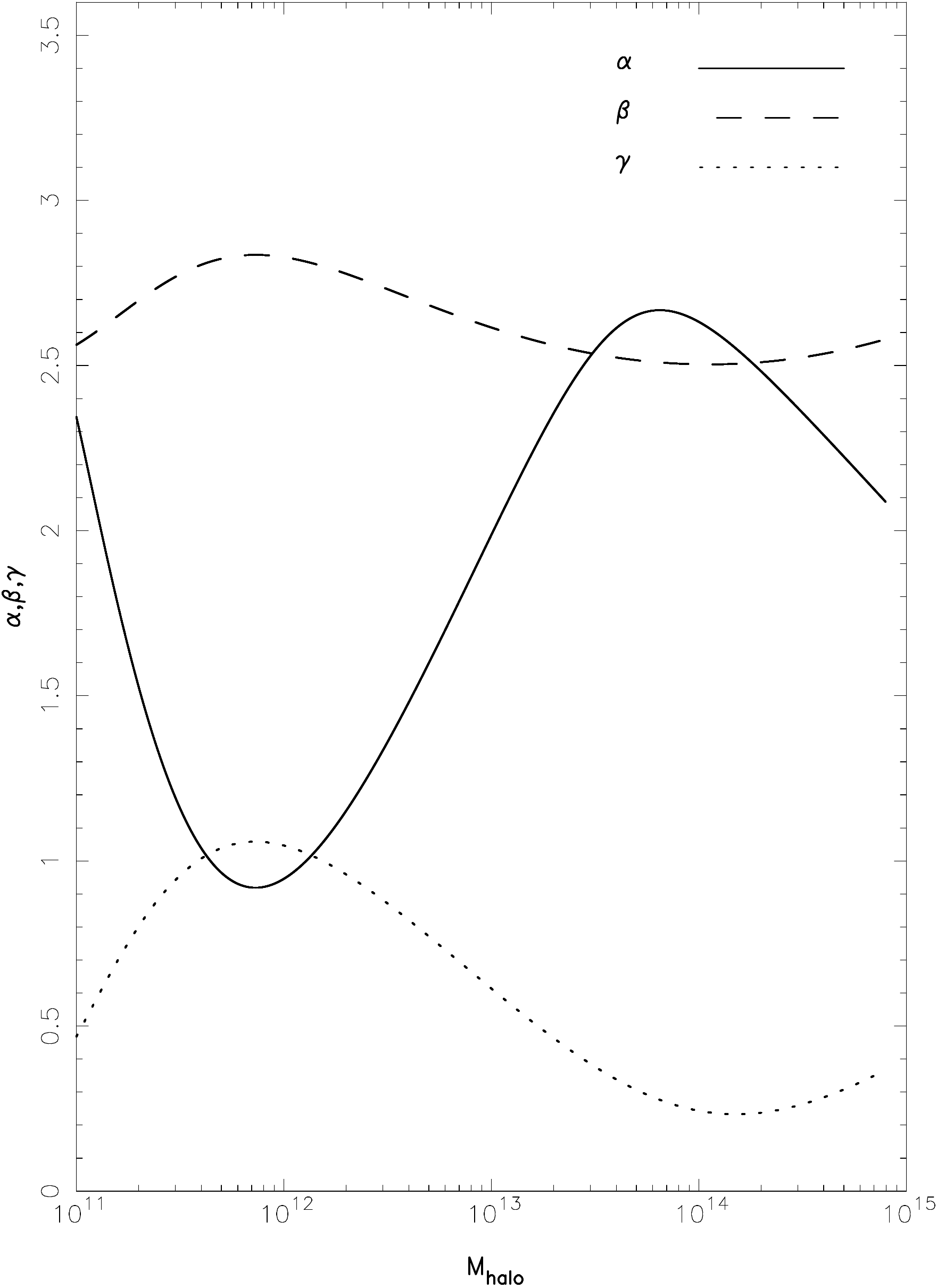}}
}
\caption{ The parameters of $\alpha$, $\beta$ and $\gamma$ as functions of 
$M_{\rm halo}$.}
\label{figure1}
\end{center}
\end{figure}

We can get some simple but important results about the lensing efficiency for 
DC14 model even before calculating the lensing probabilities. We notice that, 
from equation (\ref{eq:DC14}), NFW profile is a specific form that has 
$(\alpha, \beta, \gamma)=(1,3,1)$, 
and SIS profile is similar to  $(\alpha, \beta, \gamma)=(2,2,2)$, where in the 
latter case, $\alpha$ can be any non-zero number and we let $\alpha=2$ for 
definiteness. Therefore, it would be helpful to plot the parameters $\alpha$, 
$\beta$ and $\gamma$ versus $M_{\rm halo}$ from $10^{11}M_{\sun}$ to 
$10^{14.9}M_{\sun}$, as shown in Fig.\ref{figure1}. We 
find that the inner slope has $\gamma\sim 1$, an NFW like value, only in a 
narrow range of halo mass around $10^{12}M_{\sun}$, and it flattens for lower 
and higher mass ranges. This would result in a much lower lensing efficiency 
compared with that of NFW model and even further lower  compared with SIS 
model. 

\begin{figure}
\begin{center}
\mbox{
\hspace{-0.7cm}
\resizebox{6.7cm}{!}{\includegraphics{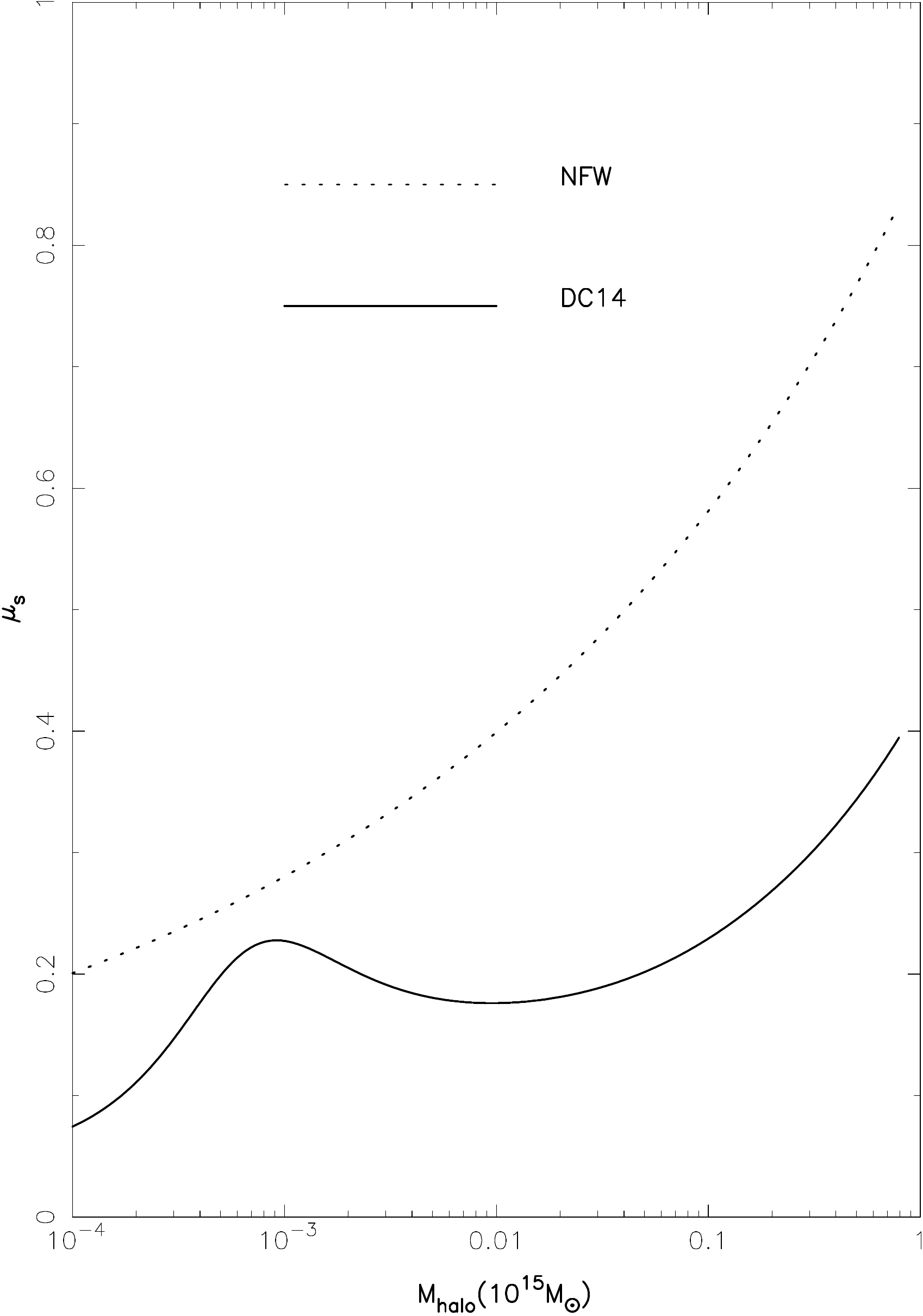}}
}
\caption{The parameter $\mu_s$ as a function of $M_{\rm halo}$ for NFW 
(dotted line) and DC14 (solid line). In both cases, the 
redshifts of source and lens are set to be 1.56 and 0.45, respectively.}
\label{figure2}
\end{center}
\end{figure}

Another parameter that helps us to understand the lensing efficiency is $\mu_s$ 
in lensing equation (\ref{lens1}). According to strong lensing theory, multiple 
images can occur only for sufficiently large values of $\mu_s$. Fig. 
\ref{figure2} shows how $\mu_s$ changes with $M_{\rm halo}$ for both DC14 and 
NFW models, with a fixed typical value $0.45$ of the lens redshift and the 
average value $1.56$ of the source redshift (\citealt{Inada12}).  We find that 
the values of $\mu_s$ for NFW model uniformly surpass that of DC14 in the whole 
range of halo mass of $(10^{11}M_{\sun}, 10^{14.9}M_{\sun})$, and the 
difference increases markedly for $M_{\rm halo}>10^{12}M_{\sun}$.  This would 
also means a obviously lower lensing efficiency for DC14 model compared with 
the NFW model.

\subsection{Schaller15 model}
\begin{figure}
\begin{center}
\mbox{
\hspace{-0.7cm}
\resizebox{6.7cm}{!}{\includegraphics{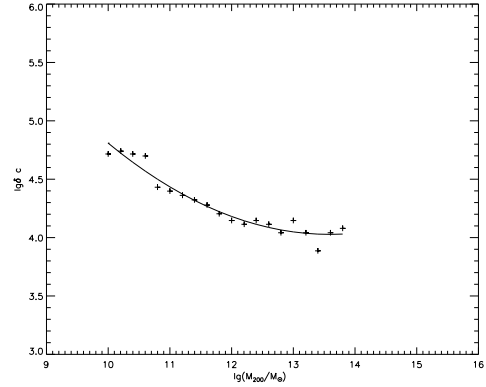}}
}
\caption{The relation between $\delta_{c}$ and $M_{200}$. }
\label{deltac}
\end{center}
\end{figure}

\begin{figure}
\begin{center}
\mbox{
\hspace{-0.7cm}
\resizebox{6.7cm}{!}{\includegraphics{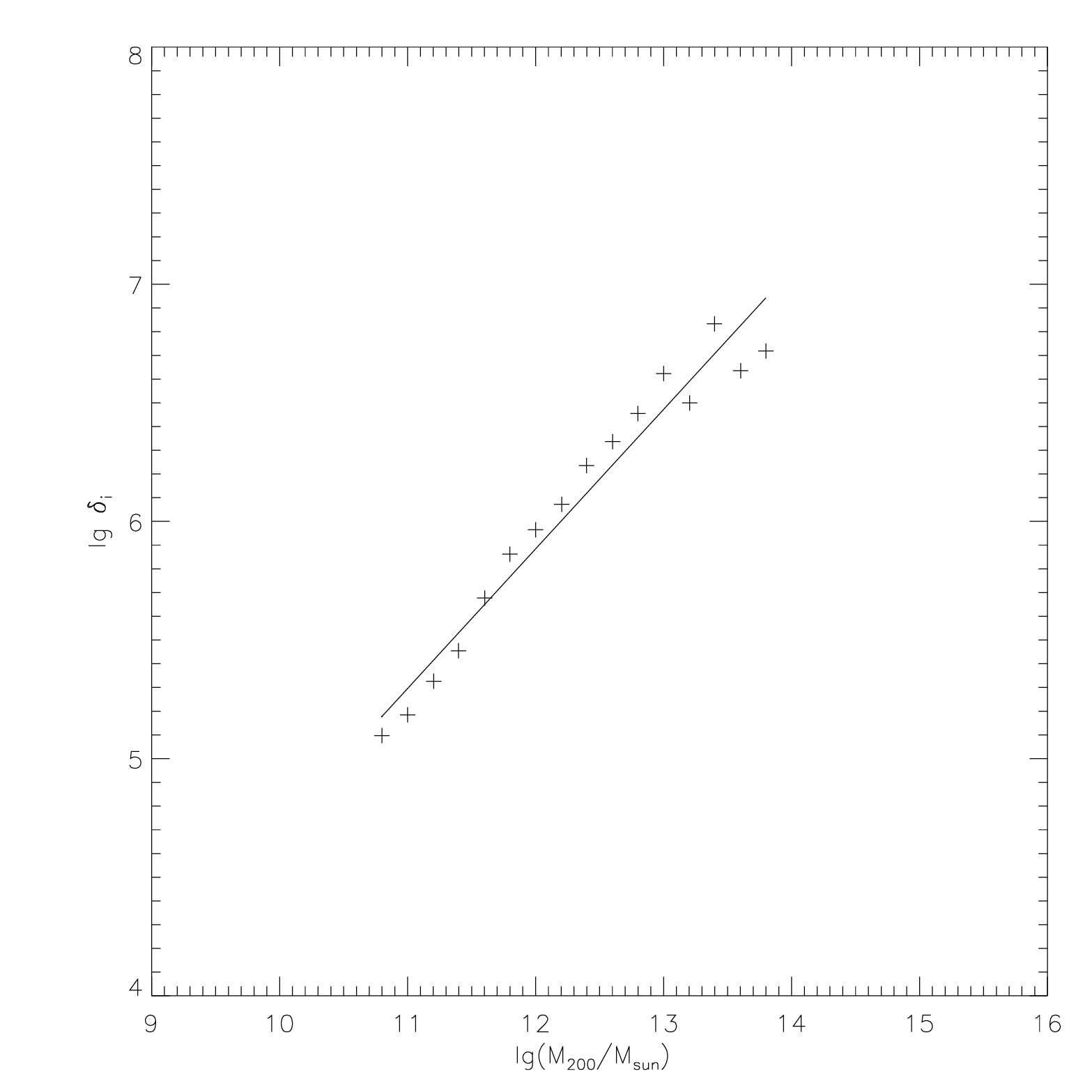}}
}
\caption{The relation between $\delta_{i}$ and $M_{200}$.}
\label{deltai}
\end{center}
\end{figure}

The Schaller15 model is derived from EAGLE simulations, the total (baryons plus 
DM) density profile consists of two terms \citep{Schaller15}
\begin{equation}
\frac{\rho(r)}{\rho_{\rm cr}}= \frac{\delta_{\rm c}}{\left(\frac{r}{r_{\rm 
s}}\right)\left(1+\frac{r}{r_{\rm 
s}}\right)^{2}}+\frac{\delta_{\rm i}}{\left(\frac{r}{r_{\rm 
i}}\right)\left[1+\left(\frac{r}{r_{\rm 
i}}\right)^{2}\right]},
\label{Schaller}
\end{equation}
where the first term is the NFW profile, and the second term is NFW-like in 
that it shares the same asymptotic behavior at small and large radii and has a 
slope of -2 at its scale radius, $r=r_{\rm i}$. We write the surface mass 
densities corresponding to the two terms as
\begin{eqnarray}
    \Sigma_1(x) = 2\rho_{\rm cr} \delta_{\rm c} r_{\rm s} V_1(x),  
\end{eqnarray}
where
\[
V_{1}(x)=\int_0^{\infty}\left(x^2+z^2\right)^{-1/2}
       \left[\left(x^2+z^2\right)^{1/2}+1\right]^{-2}dz,
\]
and 
\begin{eqnarray}
    \Sigma_2(x) = 2\rho_{\rm cr} \delta_{\rm i} r_{\rm i} V_2(x),
\end{eqnarray}
with
\[
V_{2}(x)=\int_0^{\infty}\left(x^2+z^2\right)^{-1/2}
       \left[\left(x^2+z^2\right) \frac{r_{\rm s}^{2}}{r_{\rm 
i}^{2}}+1\right]^{-1}dz.
\]
For later calculations, we need to know the 
characteristic densities $\delta_{\rm c}$, $\delta_{\rm i}$ and characteristic 
radii $r_{\rm s}$, $r_{\rm i}$ as functions of the total mass $M_{200}$, which 
reads \citep{Schaller15}
\begin{eqnarray}
 M_{200}&=&2\pi\rho_{\rm cr}\left\{2\delta_{\rm c}r^3_{\rm 
s}\left[\ln\left(1+\frac{r_{200}}{r_{\rm 
s}}\right)-\frac{r_{200}}{r_{200}+r_{\rm s}}\right]\right. \nonumber \\ 
& &\left. +\delta_{\rm i}r^3_{\rm i}\ln\left(1+\frac{r^2_{200}}{r^2_{\rm 
i}}\right)\right\}.
\label{mass}
\end{eqnarray}
In practice, however, it is impossible to derive so many parameters from the 
only equation (\ref{mass}), we thus fit $\delta_{\rm c}$ etc. to $M_{200}$ 
with the data given by \citet{Schaller15}. The fitted results are displayed in 
Fig.\ref{deltac},  Fig.\ref{deltai} and Fig.\ref{rs}. The fitting formulas are 
\begin{eqnarray}
    \delta_{\rm c}=10^{ \delta_{\rm ca}X^2+ \delta_{\rm 
cb}X+ \delta_{\rm cc}},
\end{eqnarray}
where $X=\log10 (M_{200})$ and
 $\delta_{\rm ca}=0.06$, $\delta_{\rm 
cb}=-1.65$, $\delta_{\rm cc}=15.22$;
\begin{eqnarray}
    \delta_{\rm i}=10^{0.59X-1.18};
\end{eqnarray}
\begin{eqnarray}
   r_\mathrm{s} =\frac{0.75}{10^3}(r_{\rm sa}X^3+ 
r_{\rm sb}X^2+r_{\rm sc}X+r_{\rm sd}) {\rm h}^{-1}{\rm Mpc}
\label{rs1}
\end{eqnarray}
where $r_{\rm sa}=2.08, r_{\rm sb}=-61.0, 
r_{\rm sc}=598.57,r_{\rm sd}=-1960.43$; and for $r_{\rm i}$, we adopt the 
average value $\langle r_{\rm i}\rangle=2.27$kpc \citep{Schaller15}, or
\begin{eqnarray}
    r_{\rm i}=1.70\times10^{-3}{\rm h}^{-1}{\rm Mpc}.
\end{eqnarray}
We thus obtain the lensing equation for a 
Schaller15 halo  
\begin{eqnarray}
    y = x - \mu_{\rm s} {g_{1}(x)\over x} - \mu_{\rm i} {g_{2}(x)\over x},    
\end{eqnarray}
where $y$ and $x$ are defined in the same way as for DC14 model, 
$\mu_{\rm s} ={4\rho_{\rm cr} \delta_{\rm c} r_{\rm s}/\Sigma_{\rm cr}}$, 
$\mu_{\rm i}= {4\rho_{\rm cr} \delta_{\rm i} r_{\rm i}/
\Sigma_{\rm cr}}$  and 
\begin{eqnarray}
    g_{1}(x) \equiv \int_0^x u V_{1}(u) du,
\end{eqnarray}
\begin{eqnarray}
    g_{2}(x) \equiv \int_0^x u V_{2}(u) du.
\end{eqnarray}

As indicated by \citet{Schaller15}, Schaller15 profile has two lengthscales, 
$r_{\rm s}$ and $r_{\rm i}$, where the former describes the NFW-like outer 
parts of the halo, and the latter the deviations from NFW in the inner regions. 
The second term in Eq.(\ref{Schaller}) is the inner component, which is 
characterized by two quantities, a scale radius $r_{\rm i}$ and a density 
contrast $\delta_{\rm i}$. This inner profile is an empirical model that 
describes the deviation from NFW due to the presence of stars and some 
contraction of the DM. So we expect that the lensing efficiency for Schaller15 
model should be higher than NFW.

\begin{figure}
\begin{center}
\mbox{
\hspace{-0.7cm}
\resizebox{6.7cm}{!}{\includegraphics{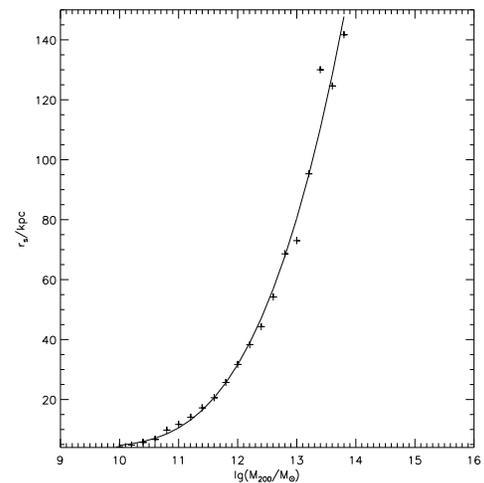}}
}
\caption{The relation between $r_{s}$ and $M_{200}$.}
\label{rs}
\end{center}
\end{figure}

\section{LENSING PROBABILITIES}
\label{sec:probab}
The quasars of redshift $z_{\mathrm{s}}$ are lensed by foreground CDM 
halos of galaxy clusters and galaxies, the lensing probability with image 
separations larger than $\Delta\theta$ is \citep{schne}
\begin{eqnarray}
P(>\Delta\theta)&=&\int^{z_{\mathrm{s}}}_0\frac{dD^\mathrm{P}_{\mathrm{L}}(z)}
{dz}dz \nonumber \\
& &\times\int^{\infty}_0\bar{n}(M,z)\sigma(M,z)B(M,z)dM ,
\label{prob1}
\end{eqnarray}
where  $D^\mathrm{P}_{\mathrm{L}}(z)$ is the
proper distance from the observer to the lens located at redshift $z$
\begin{equation}
D^\mathrm{P}_{\mathrm{L}}(z)=\frac{c}{H_0}\int^z_0\frac{dz}{(1+z)
\sqrt{\Omega_{\mathrm{m}}(1+z)^3+\Omega_{\Lambda}}},
\end{equation}
here $c$ is the speed of light in vacuum and $H_0$ is the current Hubble 
constant. We make $z_{s}=1.56$ for statistical sample SQLS (\citealt{Inada12}), 
and $z_{s}=1.27$ which is the mean value of the redshift 
distribution for quasars approximated by a Gaussian model 
\citep{Helbig99,Marlow2000,Myers03}. The physical number density $\bar{n}(M,z)$ 
of virialized DM halos of 
masses between $M$ and $M+dM$ is related to the comoving number density 
$n(M,z)$ 
by $\bar{n}(M,z)=n(M,z)(1+z)^3$; the latter was originally given by 
\cite{press74}, and the improved version is \citep{Sheth99}
\begin{equation}
n(M,z)dM=\frac{\rho_\mathrm{crit}}{M}f(M,z)dM,
\end{equation}
where 
\begin{equation}
f(M,z)=-\sqrt{\frac{2}{\pi}}\frac{\delta_c(z)}{M\Delta}
\frac{d\ln\Delta}{d\ln
M}\exp\left[-\frac{\delta_c^2(z)}{2\Delta^2} 
\label{ps}\right]
\end{equation}
is PS mass function. In Eq.(\ref{ps}) above, $\Delta^2(M)$ is the present 
variance of the fluctuations in a sphere containing a mean mass $M$,
\begin{equation}
\Delta^2(M)=\frac{1}{2\pi^2}\int^{\infty}_0P(k)
W^2(kr_{\mathrm{M}})k^2dk,
\end{equation}
where $P(k)$ is the power spectrum of density fluctuations, 
$W(kr_{\mathrm{M}})$ is the Fourier transformation of a top-hat window function
\begin{equation}
W(kr_{\mathrm{M}})=3\left[\frac{\sin(kr_{\mathrm{M}})}
{(kr_{\mathrm{M}})^3}-\frac{\cos(kr_{\mathrm{M}})}
{(kr_{\mathrm{M}})^2}\right],
\end{equation}
and
\begin{equation}
r_{\mathrm{M}}=\left(\frac{3M}{4\pi\rho_0}\right)^{1/3}.
\end{equation}
In Eq.(\ref{ps}), $\delta_c(z)$ is the over density threshold for spherical 
collapse by redshift $z$ (\citealt{nfw97}):
\begin{equation}
\delta_{c}(z)=\frac{1.68}{D(z)},
\end{equation}
where $D(z)$ is the linear growth function of density perturbation 
\citet{carroll-Press}
\begin{equation}
D(z)=\frac{g(\Omega(z))}{g(\Omega_{\mathrm{m}})(1+z)},
\end{equation}
in which
\begin{equation}
g(x)=\frac{5}{2}x\left(\frac{1}{70}+\frac{209x}{140}
-\frac{x^2}{140}+x^{4/7}\right)^{-1},
\end{equation}
and
\begin{equation}
\Omega(z)=\frac{\Omega_{\mathrm{m}}(1+z)^3}
{1-\Omega_{\mathrm{m}}+\Omega_{\mathrm{m}}(1+z)^3}.
\end{equation}
We use the fitting formulae for CDM power spectrum $P(k)$ given by \citet{Eisen}
\begin{equation}
P(k)=AkT^2(k),
\end{equation}
where $A$ is the amplitude normalized to
$\sigma_8=\Delta(r_{\mathrm{M}}=8h^{-1}\mathrm{Mpc})=0.8$, and
\begin{equation}
T=\frac{L}{L+Cq^2_{\mathrm{eff}}},
\end{equation}
with
\begin{equation}
L\equiv\ln(e+1.84q_{\mathrm{eff}}),
\end{equation}
\begin{equation}
q_{\mathrm{eff}}\equiv\frac{k}{\Omega_{\mathrm{m}}h^2{\mathrm{Mpc^{-1}}}},
\end{equation}
\begin{equation}
C\equiv14.4+\frac{325}{1+60.5q_{\mathrm{eff}}^{1.11}}.
\end{equation}
\noindent The cross-section is
\begin{equation}
\sigma(M, z)=\pi
y_{\mathrm{cr}}^2r_{\mathrm{s}}^2\vartheta(M-M_{\mathrm{min}}),
\end{equation}
where $y_\mathrm{cr}$ is the maximum value of $y$, the reduced position of a
source, such that when $y<y_\mathrm{cr}$ multiple images can occur;  
$\vartheta(x)$ is a step function, and $M_{\mathrm{min}}$ is determined 
by the lower limit of image separation
\begin{equation}
\Delta\theta=\frac{r_{\mathrm{s}}\Delta
x}{D_\mathrm{L}}\approx\frac{2x_0r_{\mathrm{s}}}{D_\mathrm{L}}
\label{dtheta}
\end{equation}
and Eq.(\ref{rs}) for DC14 model as
\begin{eqnarray}
M_{\mathrm{min}}^{\mathrm{DC14}}&=&8.927\times10^{-8}M_{15} \nonumber \\
& &\times\left(\Omega_{\mathrm{m}}(1+z)^3+\Omega_{\Lambda}\right)
\left(\frac{c_1D_\mathrm{L}\Delta\theta}{x_0}\right)^3,
\end{eqnarray}
and Eq.(\ref{rs1}) for Schaller15 model as 
\begin{eqnarray}
M_{\mathrm{min}}^{\mathrm{Schaller15}}&=10^{m_{\rm a} 
Y^3+m_{\rm b}Y^2+m_{\rm c}Y+M_{\rm d}-15},
\end{eqnarray}
where $Y=\frac{10^3}{0.75}\times 
r_{\rm s}$, $m_{\rm a}=3.62\times10^{-6},m_{\rm 
b}=-0.001, m_{\rm c}=0.09, m_{\rm d}=9.92$. Note that, for Schaller15 model, we 
define $M=M_{200}/10^{15}M_{\sun}$. In Eq.(\ref{dtheta}), we have approximated 
the image separation $\Delta x$ to be $2x_0$, where $x_0$ is the positive zero 
position of function $y(x)$.

The  magnification bias $B(M,z)$ should be calculated by considering the actual 
flux ratio and differential luminosity of quasar sources (e.g., 
\citealt{Oguri08,YC09}), however, since we investigate only the order of 
magnitudes of lensing probabilities, we adopt a simple model (\citealt{li02}):
$B\approx 2.2A_{m}^{1.1}$, with 
$A_m=D_\mathrm{L}\Delta\theta/(r_sy_\mathrm{cr})$.

\begin{figure}
\begin{center}
\mbox{
\hspace{-0.7cm}
\resizebox{6.7cm}{!}{\includegraphics{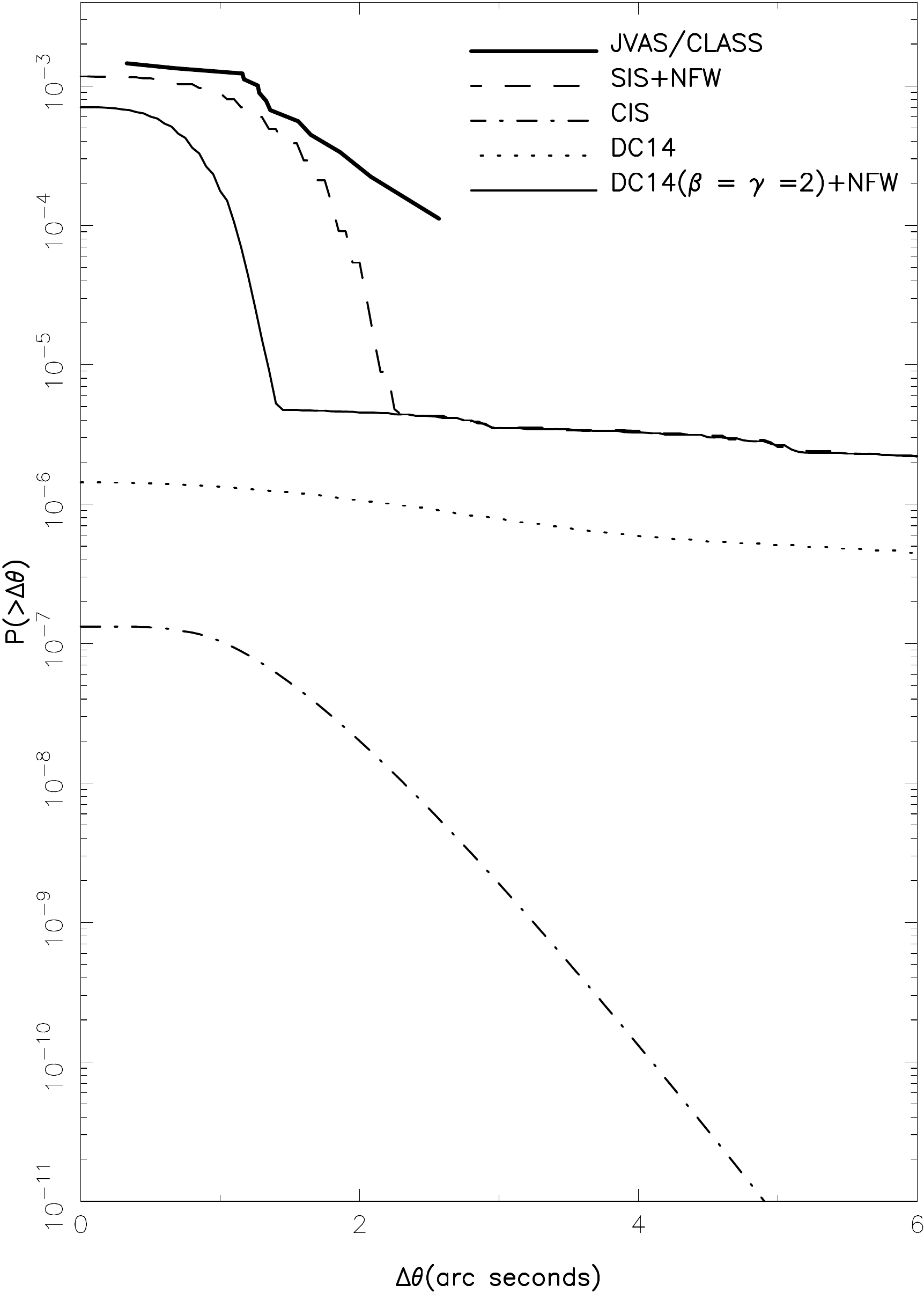}}
}
\caption{Lensing probabilities with image separations larger than 
$\Delta\theta$: observations of the combined JVAS/CLASS sample (thick 
histogram), and the predictions for the models of SIS+NFW (dashed line),  CIS 
(dot-dashed), DC14($\beta=\gamma=2)$+NFW (solid line), and DC14 (dotted 
line). Predicted lensing probabilities are calculated with $z_s=1.27$.}
\label{figure6}
\end{center}
\end{figure}

We first present, in Fig.~\ref{figure6}, the lensing probabilities predicted 
by Eq.(\ref{prob1}) with the survey results of JVAS/CLASS, which is a subset 
of 8958 sources from the combined JVAS/CLASS survey that forms a 
well-defined statistical sample containing 13 multiply imaged sources suitable 
for analysis of the lens statistics \citep{Myers03,Browne03,King99}. The 
lensing probability for DC14 model (the dotted line) is much lower than  the
observations, which verifies our previous predictions simply based on the inner 
slope and parameter $\mu_s$. Also shown in Fig.~\ref{figure6} is the well known 
two-population SIS+NFW model, which has long been used as a
standard model in strong lensing statistics (\citealt{Sarbu01, 
li02,Chen03a,Chen03b,Chen04a,Chen04b,Zhang04}). In this model, SIS is used for 
lensing galaxies (mostly giant ellipticals) and NFW  for lensing clusters of 
galaxies \citep{Boldrin12}, and the transition occurs at 
$M_\mathrm{halo}\sim 10^{13}M_{\sun}$.  
We can conclude that the SIS+NFW model fit the observations reasonably well, in 
the sense that the SIS predictions fit the small image separations well, whilst 
the NFW predictions are below the upper limit put by JVAS/CLASS survey  for 
$6''\leq\Delta\theta\leq 15''$ \citep{li02}.  We know that the 
density profile for SIS model is $\rho_\mathrm{SIS}(r)=\sigma^2_v/(2\pi Gr^2)$, 
where $\sigma_v$ is the velocity dispersion; alternatively, if we set 
$\beta=\gamma=2$ in 
Eq.~(\ref{eq:DC14}) we have $\rho(r)=\rho_{s}r^2_s/r^2$. They are both 
proportional to $1/r^2$, and differ only in a constant. The latter case is 
denoted as ``DC14($\beta=\gamma=2$)+NFW'' (the dot-dash line) in 
Fig.~\ref{figure6}, which is approximately equivalent to SIS+NFW model. This 
reflects a very important fact about strong lensing statistics that we have
repeatedly emphasized: the inner slope of density profile for 
lensing halos is the most important factor compared with others (e.g, 
shapes and substructures). The lensing probabilities for CIS model have been 
investigated in detail \citep{CM10}, we paste the line (dot-dash) in 
Fig.~\ref{figure6}. We find that the lensing probabilities for DC14  are lower 
than NFW but higher than the CIS model, which can be explained by the 
steepening tendency of the inner slope of the DC14 profile when the halo mass 
increases towards $\sim 10^{12}M_{\sun}$, as displayed in Fig.~\ref{figure1}.  

\begin{figure}
\begin{center}
\mbox{
\hspace{-0.7cm}
\resizebox{6.7cm}{!}{\includegraphics{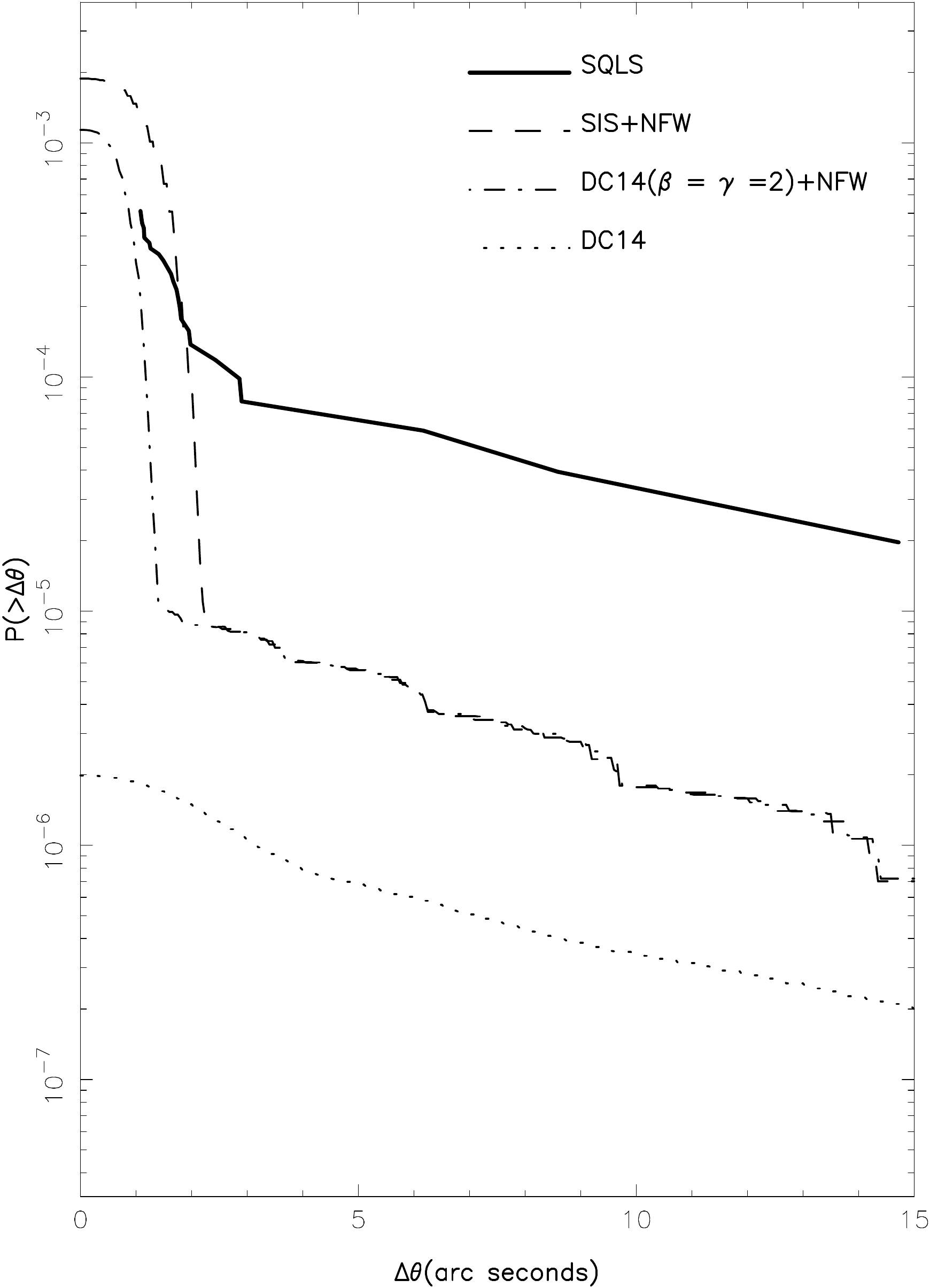}}
}
\caption{Lensing probabilities with separation larger 
than $\Delta\theta$: observations for SQLS sample (thick histogram), and the
predictions for the models of SIS +NFW (dashed line), DC14($\beta$ = 
$\gamma$ =2) + NFW (dot-dashed line) and DC14 (dotted line). Predicted lensing 
probabilities are calculated with $z_s=1.56$.}
\label{figure7}
\end{center}
\end{figure}

We also compare the lensing probabilities predicted with Eq.(\ref{prob1}) for 
various density profiles with the most recent observations of SQLS in 
Fig.~\ref{figure7}. The statistical sample for SQLS (\citealt{Inada12}) 
consists of 26 quasar lenses selected from 50836 source quasars in the 
redshift range  0.6 $<$ z $<$ 2.2 with Galactic extinction corrected 
\citep{Schlegel98} magnitudes brighter than 
$i$ = 19.1. Note that the predicted lensing probabilities for each model in 
Fig.~\ref{figure7} are obviously higher than their counterparts displayed in 
Fig.~\ref{figure6} due to the different redshifts $z_s$ of quasars we have 
chosen, $z_s=1.56$ for SQLS and $1.27$ for JVAS/CLASS.   We find from 
Fig.~\ref{figure7} that SIS profile can still match the SQLS observations well, 
whilst NFW predicts the lensing probabilities that are about an order of 
magnitude lower than the observations \citep{Giocoli16}. The 
usually employed standard model 
SIS+NFW breaks down for large image separations. As pointed out previously, 
something like the ellipticity and substructure, which deviate from the 
spherical and smooth NFW model, cannot compensate for the large discrepancy,  
we thus tend to believe that a steeper inner slope than NFW may achieve the 
large image separation observations \citep{CM10}. 

\begin{figure}
\begin{center}
\mbox{
\hspace{-0.7cm}
\resizebox{6.7cm}{!}{\includegraphics{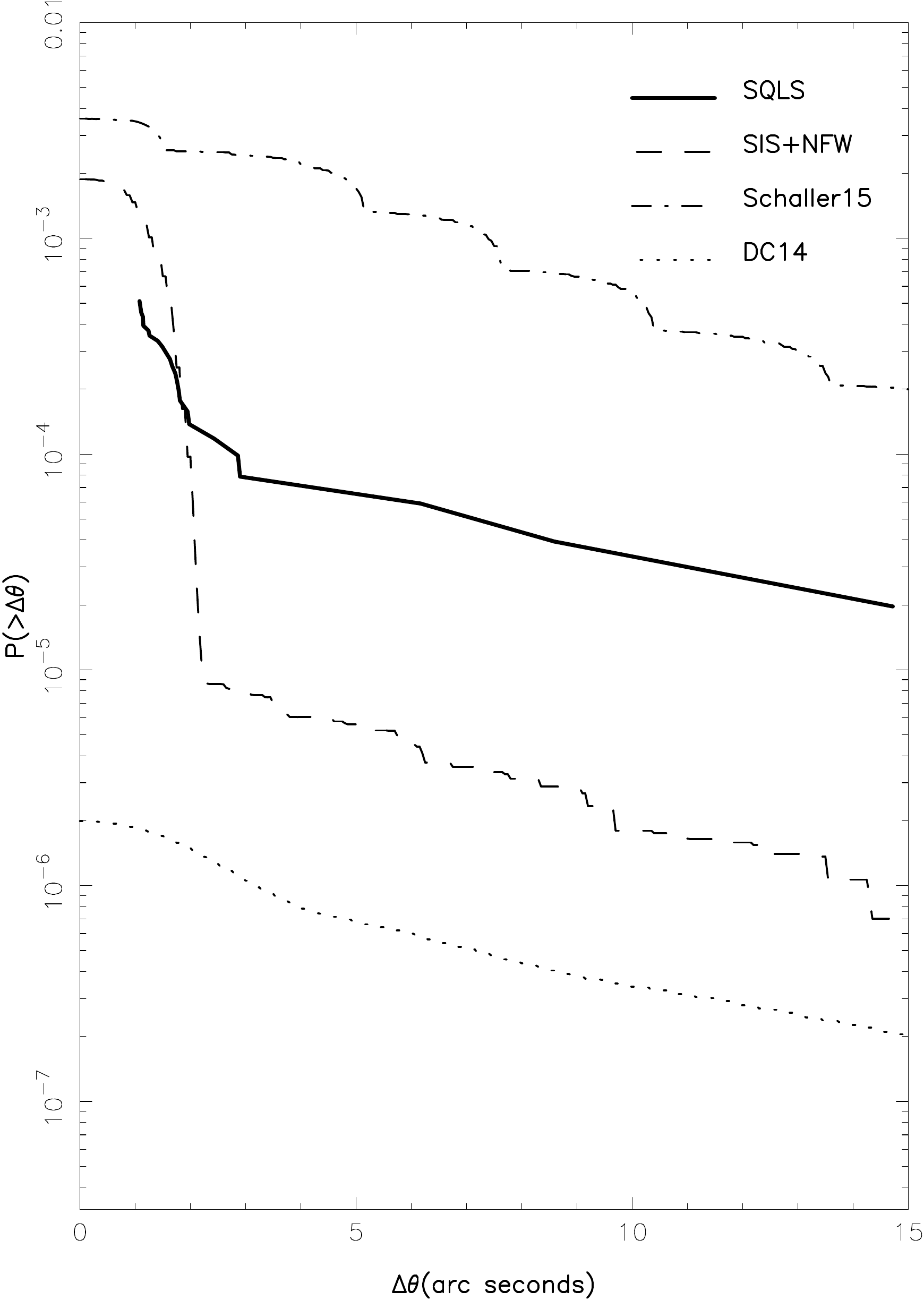}}
}
\caption{Lensing probabilities with separations larger 
than $\Delta\theta$: observations for SQLS sample (thick histogram), and the
predictions for the models of SIS +NFW (dashed line), Schaller15 
(dot-dashed line) and DC14 (dotted line). Predicted lensing 
probabilities are calculated with $z_s=1.56$.}
\label{figure8}
\end{center}
\end{figure}

The predicted lensing probabilities for Schaller15 model (dot-dashed line) are 
shown in Fig.~\ref{figure8}, together with the observations for SQLS sample 
(thick histogram), the predictions for the models of SIS +NFW (dashed line) and 
DC14 (dotted line). Surprisingly, we find that Schaller15 model predicts too 
many lenses compared with SQLS observations and all other models. 

\section{Discussions and Conclusions }
\label{sec:conc}
We have calculated the lensing probabilities with image separations larger than 
$\Delta\theta$ for DC14 model and Schaller15 model, and compared the 
results with observations 
and other models. As expected, the lensing efficiency for DC14 is much less 
than SIS (which fit the observations for galactic lenses quite well), and even 
less than NFW. The reason is that the lensing efficiency is very sensitive to 
the inner slope of the density profile of the lens halos, which actually 
dominates the predictions. Despite that the inner slope $\gamma$ of DC14 
profile approaches 1 (which is NFW like) when the halo mass increases towards 
$10^{12}M_{\sun}$, it decreases dramatically after that mass as shown in 
Fig.~\ref{figure1}. We know that the DC14 profile is fitted from the 
simulations 
which are confined to $M_\mathrm{halo}\leq 10^{12}M_{\sun}$, and thus should be 
valid only in this range; however, there are no evidences arising from the 
treatments of astrophysical processes for the simulations show us that we 
cannot extrapolate to larger halos. What is more important about DC14 profile 
is that it is very far from the SIS like density profile in the galactic mass 
range, namely around $M_\mathrm{halo}\sim 10^{12}M_{\sun}$, which is required 
to explain strong lensing observations. This phenomena is, in fact, genuine in 
the literature: up to now, all the simulations claimed to have explained 
reasonably the observations 
of rotation curves fail to explain the observations of strong lensing, whatever 
the valid halo mass ranges declared. For example, one possible solution to the 
cusp-core problem is the turbulence driven by stellar feedback during galaxy 
formation \citep{MCW06,MWC08}, which leads to a final halo file with a finite 
core radius for all galaxies, including giant ellipticals. Such a situation is 
consistent with essentially all observations of rotation curves 
\citep{McGaugh07}, but contradicts with strong lensing observations 
\citep{CM10}. 

One may argue that, disk galaxies from which we observe the rotation curves 
and the  giant ellipticals which dominate strong lensing phenomena, are of very 
different galaxy types and form in very different environments and histories. In 
the spirit of hierarchical CDM structure formation paradigm, however, they are 
formed from the same initial conditions (gas mixes with dark matter) and 
undergo the same subsequent hierarchical sequences, and thus cannot form 
separately and independently. Clearly, any valuable and theoretically 
significant predictions of the properties of galaxies should be of those for 
any simulations that cover the mass range from dwarf galaxies to giant 
ellipticals, and should have the volume 
size large enough to include the statistically well-defined samples of 
galaxies. This is necessary to ensure the hierarchical
galaxy formation theory be faithfully, coherently and self-consistently 
implemented in the $\Lambda$CDM paradigm. Therefore, for any simulations, 
whatever the manners are assumed in which baryon effects are modeled to modify 
the initially pure DM halo profiles, it is difficult, if not 
impossible, to identify the final galaxies hosting in the center of DM halos 
only with disk galaxies, in particular when the halo mass is as large as 
$10^{12}M_{\sun}$. That is, when the redshift $z\sim 0$,  there should 
exist other galaxy types apart from disk galaxies. We thus emphasize that, it 
is meaningful  for DC14 and any other similar density profiles to be tested 
against rotation curves, only if we assume that the halo mass density profiles 
are unconcerned with morphological types of the hosted galaxies. Accordingly, 
the density profiles can also, and should be, tested by other available 
observations, in particular by the observations of strong lensing 
\citep{Chen05, LC09, CM10}. 

In practice, however, limited by the computer capabilities, more details about 
the inner structure of each halo need higher resolutions (i.e., smaller 
particle mass) which would strongly restrict the sample size under 
considerations. Consequently, simulations can only be designed to tackle a 
certain specific problem (usually determined by observations). For example, 
disk galaxies and giant ellipticals are often simulated independently, usually 
among very different communities. The baryons have two opposite 
effects on the central mass density of DM halos. While stellar feedback and 
dynamical friction can induce expansion of the DM halo and produce a core, the 
adiabatic contractions can steepen central density to the SIS type 
\citep{BFFP86,GKKN04,GFS06}. The uncertainties of the parameters appear in  
different models for the baryon effects  allow us to calibrate the parameters 
with observations. This inevitably leads to the simulation results which are 
strongly observation-dependent. It is thus no surprise that the baryon 
processes modeled for simulations that can produce the CIS profile cannot 
naturally proceed to produce SIS (\citealt{CM10}).

The most recent seemingly comprehensive simulations (e.g., \citealt{PEF09, 
Remus13,Schaller15}) further confirm our opinions mentioned above: a treatment 
of baron effects for one aspect of observations cannot describe another. 
Schaller15 profile has no core, however, the rotation curves of the simulated 
halos are in excellent agreement with observational data \citep{Reyes11} for 
galaxies with stellar mass ranging from $10^9M_{\sun}$ to $5\times 
10^{11}M_{\sun}$, corresponding to the total halo mass ranging from 
$10^{11}M_{\sun}$ to 
$10^{13}M_{\sun}$ \citep{Schaller15}. This is compatible with DC14 model  
in the sense that, for DC14, the cores exist only for the low mass halos 
and the profile is NFW like when the halo mass approaches $10^{12}M_{\sun}$. 
For halos with mass $\gsim 10^{12}M_{\sun}$, DC14 model has no data, our 
extrapolation predicts too low lensing efficiencies. For Schaller15 halos, 
however, the central regions of halos with mass  $\gsim 10^{12}M_{\sun}$ are 
dominated by the stellar component \citep{Schaller15}. The presence of these 
baryons causes a contraction of the halos and thus enhances the density of DM 
in this regions. Unfortunately, the over-predicted lensing efficiencies 
mean that the baryon effects on DM suggested by Schaller15 model cannot be 
true.

We conclude that, it is difficult for current simulations to 
reconcile the DM 
distributions derived from the observations of rotation curves and that from 
strong lensing. In the context of $\Lambda$CDM cosmology, if baryon effects, in 
the computer simulations, are treated specifically to fit some specific 
observations without considering others,  results need to be more aware of the 
systematics and the limitations of both theory and observations.

\section*{Acknowledgements}
We thank the referee for providing constructive comments and help in improving 
the contents of this paper. This work was supported by the 
National Natural Science Foundation of China 
(Grant 11073023). RL is supported by Youth Innovation Promotion Association of 
CAS and Youth Science Funding of NAOC.

\setlength{\bibhang}{2.0em}


\begin{thebibliography}{}
\setlength{\itemindent}{-2.5em}

\bibitem[\protect\citeauthoryear{Angulo et al.} {2012}]{Angulo12}
Angulo R. E., Springel V., White S. D. M., Jenkins A., Baugh C. M., Frenk C. 
S., MNRAS, 426, 2046


\bibitem[\protect\citeauthoryear{Bartelmann et al.} {1998}]{Bartelmann98}
Bartelmann M. et al., A\&A, 330,1 


\bibitem[\protect\citeauthoryear{Begeman et al.} {1991}]{Begeman91} 
Begeman K. G., Broeils A. H., Sanders R. H., 1991, MNRAS, 249, 523
\bibitem[\protect\citeauthoryear{Berlind \& Weinberg} {2002}]{BW02} Berlind
A.~A., Weinberg D. H., 2002,  ApJ, 575, 587
\bibitem[\protect\citeauthoryear{Blumenthal et al.}{1986}]{BFFP86}Blumenthal G. 
R., Faber S. M., Fores R., Primack J. R., 1986, ApJ, 301, 27


\bibitem[\protect\citeauthoryear{Boldrin et al.}{2012}]{Boldrin12} Boldrin M., 
Giocoli C., Meneghetti M., Moscardini L., 2012, MNRAS, 427, 3134



\bibitem[\protect\citeauthoryear{Bonamigo et al.}{2015}]{Bonamigo15}Bonamigo 
M., Despali G., Limousin M., Angulo R., Giocoli C., Soucail G., 2015, MNRAS, 
449, 3171



\bibitem[\protect\citeauthoryear{Bond et al.}{1991}]{BCEK91}Bond J. R., Cole 
S., Efstathiou G., Kaiser N. 1991, ApJ, 379, 440
\bibitem[\protect\citeauthoryear{Bower}{1991}]{Bower91}Bower R. G., 1991, 
MNRAS, 248, 332
\bibitem[\protect\citeauthoryear{Browne et al.}{2003}]{Browne03}
Browne I. W. A. et al., 2003, MNRAS, 341, 13

\bibitem[\protect\citeauthoryear{Broadhurst \& Barkana} {2008}] {Broadhurst08}
Broadhurst T. J., Barkana R., 2008, MNRAS, 390, 1647


\bibitem[\protect\citeauthoryear{Bullock et al.} {2002}]{Bullock02} Bullock J.
S., Wechsler R. H., Somerville R. S., 2002, MNRAS, 329, 246
\bibitem[\protect\citeauthoryear{Carroll \& Press}{1992}]{carroll-Press}
Carroll S. M., Press W. H., 1992, ARAA, 30, 499
\bibitem[\protect\citeauthoryear{Chen} {2003a}]{Chen03a}
Chen D. -M., 2003a, A\&A, 397, 415
\bibitem[\protect\citeauthoryear{Chen} {2003b}]{Chen03b}
Chen D. -M., 2003b, ApJ, 587, L55
\bibitem[\protect\citeauthoryear{Chen} {2004a}]{Chen04a}
Chen D. -M., 2004a, A\&A, 418, 387
\bibitem[\protect\citeauthoryear{Chen} {2004b}]{Chen04b}
Chen D. -M., 2004b, Chinese J. Astron. Astrophys., 4, 118
\bibitem[\protect\citeauthoryear{Chen}{2005}]{Chen05}Chen D.-M., 2005, ApJ, 
629, 23
\bibitem[\protect\citeauthoryear{Chen \& McGaugh}{2010}]{CM10}Chen D. -M., 
McGaugh S., 2010, RAA, 10, 1215
\bibitem[\protect\citeauthoryear{Commerçon, Debout \& 
Teyssier}{2014}]{Commercon14}	
Commerçon B., Debout V., Teyssier R., 2014, A\&A, 563, 11

\bibitem[\protect\citeauthoryear{Conroy et al.} {2006}]{Con06} Conroy C.,
Wechsler R. H., Kravtsov A. V., 2006, ApJ, 647, 201


\bibitem[\protect\citeauthoryear{Courtin et al.} {2010}]{Courtin10} 


\bibitem[\protect\citeauthoryear{Crocce et al.} {2010}]{Crocce10} Crocce M., 
Fosalba P., Castander F. J., Gaztanaga E., 2010, MNRAS, 403, 1353



\bibitem[\protect\citeauthoryear{de Blok et al.} {2001}]{deBlok01} de Blok W. 
J. G., McGaugh S. S., Bosma A., Rubin V. C.,  
2001, ApJ, 552, L23
\bibitem[\protect\citeauthoryear{de Blok et al.} {2008}]{deBlok08} de Blok W. 
J. G., Walter F., Brinks E., Trachternach C., Oh S.-H., Kennicutt R. C., Jr, 
2008, AJ, 136, 2648
\bibitem[\protect\citeauthoryear{de Blok}{2010}]{deBlok10}de Blok W. J. G., 
2010, Advances in Astronomy, Volume 2010, Article ID 789293

\bibitem[\protect\citeauthoryear{Despali et al.}{2014}]{Despali14} Despali G., 
Giocoli C., Tormen G., 2014, MNRAS, 443, 3208


\bibitem[\protect\citeauthoryear{Despali et al.}{2016}]{Despali16} Despali G., 
Giocoli C., Angulo R. E., Tormen G., Sheth R. K., Baso G., Moscardini L., 2016, 
MNRAS, 456, 2486


\bibitem[\protect\citeauthoryear{Despali et al.}{2017}]{Despali17} Despali G., 
Giocoli C., Bonamigo M., Limousin M., Tormen G., 2017, MNRAS, 466, 181

\bibitem[\protect\citeauthoryear{Di Cintio et al.} {2014}]{Cintio14} Di Cintio
A., Brook C.~B., {Dutton} A.~A., Maccio A.~V., Stinson G.~S., Knebe A., 2014,
MNRAS, 441, 2986
\bibitem[\protect\citeauthoryear{Eisenstein \& Hu}{1999}]{Eisen} Eisenstein D. 
J., Hu W., 1999, ApJ, 511, 5
\bibitem[\protect\citeauthoryear{Frenk \& White}{2012}]{FW12}Frenk C. S.,
White S. D. M., 2012, Annalen der Physik, 524, 507
\bibitem[\protect\citeauthoryear{Gentile et al.} {2004}]{Gentile04} Gentile G., 
Salucci P., Klein U., Vergani D., Kalberla P., 2004, MNRAS, 351, 903 



\bibitem[\protect\citeauthoryear{Giocoli et al.} {2012}]{Giocoli12} Giocoli C., 
Meneghetti M., Bartelmann M., Moscardini L., Boldrin M., MNRAS, 421, 3343


\bibitem[\protect\citeauthoryear{Giocoli et al.} {2016}]{Giocoli16} Giocoli C., 
Bonamigo M., Limousin M., Meneghetti M., Moscardini L., Angulo R. E., Despali 
G., Jullo E., 2016, MNRAS, 462, 167



\bibitem[\protect\citeauthoryear{Gnedin et al.}{2004}]{GKKN04}Gnedin O. Y., 
Kravtsov A. V., Klypin A. A., Nagai D., 2004, ApJ, 616, 16
\bibitem[\protect\citeauthoryear{Gott \& Gunn} {1974}]{Gott74} Gott J. ~R., 
Gunn J.~E., 1974, ApJ, 190, L105
\bibitem[\protect\citeauthoryear{Guo et al.} {2010}]{guo10}
Guo Q., White S., Li C., Boylan-Kolchin M., 2010, MNRAS, 404, 1111
\bibitem[\protect\citeauthoryear{Gustafsson, Fairbairn \& 
Sommer-Larsen}{2006}]{GFS06}Gustafsson M., Fairbairn M., Sommer-Larsen J., 
2006, Phys. Rev. D, 74, 123522
\bibitem[\protect\citeauthoryear{Helbig et al.} {1999}] {Helbig99} Helbig P., 
Marlow D., Quast R., Wilkinson P. N., Browne I. W. A., Koopmans L. V. 
E., 1999, A\&AS, 136, 297 

\bibitem[\protect\citeauthoryear{Hennawi et al.} {2007}] {Hennawi07}
Hennawi J. F., Dalal N., Bode P., Ostriker J. P., 2007, ApJ, 654, 714



\bibitem[\protect\citeauthoryear{Inada et al.}{2012}]{Inada12}Inada N. et 
al., 2012, AJ, 143, 119 



\bibitem[\protect\citeauthoryear{Jenkins et al.}{2001}]{Jenkins01}Jenkins A., 
Frenk C. S., White S. D. M., Colberg J. M., Cole S., Evrard A. E., Couchman H. 
M. P., Yoshida N., 2001, MNRAS, 321, 372


\bibitem[\protect\citeauthoryear{Jing et al.}{2002}]{Jing02} Jing Y. P., 
Suto Y., 2002, ApJ, 574, 538


\bibitem[\protect\citeauthoryear{Lacey \& Cole}{1993}]{LC93}Lacey C., Cole 
S., 1993, MNRAS, 262, 627

\bibitem[\protect\citeauthoryear{Katsianis et al.}{2017}]{Katsianis17}	
Katsianis A., Tescari E., Blanc G., Sargent M., 2017, MNRAS, 464, 4977
\bibitem[\protect\citeauthoryear{Katz et al.} {2016}] {Katz16}
 Katz H., Lelli F., McGaugh S. S., Di Cintio A., Brook C. B., Schombert J. M., 
 2017, MNRAS, 466, 1648  


\bibitem[\protect\citeauthoryear{Kauffmann \& White}{1993}]{KW93}Kauffmann G., 
 White S. D. M., 1993, MNRAS, 261, 921

\bibitem[\protect\citeauthoryear{King et al.}{1999}]{King99}
King L. J., Browne I. W. A., Marlow D. R., Patnaik A. R.,
Wilkinson P. N., 1999, MNRAS, 307, 255

\bibitem[\protect\citeauthoryear{Kravtsov \& Klypin} {1999}]{KK99} Kravtsov A.,
 Klypin A., 1999, ApJ, 520, 437

\bibitem[\protect\citeauthoryear{Kuzio de Naray et al.} {2008}]{Kuzio08} 
Kuzio de Naray R., McGaugh S. S., de Blok W. J. G., 2008, ApJ, 676, 920

\bibitem[\protect\citeauthoryear{Kuzio de Naray et al.} {2006}]{Kuzio06} 
Kuzio de Naray R., McGaugh S. S., de Blok W. J. G., Bosma A., 2006, ApJS, 165, 
461
\bibitem[\protect\citeauthoryear{Kuzio de Naray et al.} {2009}]{Kuzio09} 
Kuzio de Naray R., McGaugh S. S., Mihos J. C., 2009, ApJ, 692, 1321

\bibitem[\protect\citeauthoryear{Li \&Ostriker} {2002}]{li02} Li L. -X., 
Ostriker J. P., 2002, ApJ, 566, 652
\bibitem[\protect\citeauthoryear{Li \& Chen}{2009}]{LC09}Li N., Chen 
D.-M., 2009, RAA, 9, 1173
\bibitem[\protect\citeauthoryear{Marlow et al.} {2000}] {Marlow2000} 
Marlow D. R., Rusin D., Jackson N., Wilkinson P. N., Browne I. W. A.,  
Koopmans L., 2000, AJ, 119, 2629  

\bibitem[\protect\citeauthoryear{Meneghetti et al.} {2001}] {Meneghetti01}
Meneghetti M. et al., 2001, MNRAS, 325, 435

\bibitem[\protect\citeauthoryear{Meneghetti et al.} {2003}] {Meneghetti03}
Meneghetti M., Bartelmann M., Moscardini L., 2003, MNRAS, 340, 105


\bibitem[\protect\citeauthoryear{Mashchenko, Couchman \& 
Wadsley}{2006}]{MCW06}Mashchenko S., Couchman H. M. P., Wadsley J., 2006, 
Nature, 442, 539
\bibitem[\protect\citeauthoryear{Mashchenko, Wadsley \& 
Couchman}{2008}]{MWC08}Mashchenko S., Wadsley J., Couchman H. M. P., 2008, 
Science, 319, 174
\bibitem[\protect\citeauthoryear{McGaugh et al.}{2007}]{McGaugh07} 
McGaugh S., de Blok W. J. G., Schombert J. M., Kuzio de Naray R., Kim J. H., 
2007, ApJ, 659, 149 
\bibitem[\protect\citeauthoryear{Myers et al.} {2003}]{Myers03}
Myers S. T. et al., 2003, MNRAS, 341, 1
\bibitem[\protect\citeauthoryear{Neyrinck et al.} {2004}]{Ney04} Neyrinck M.
C., Hamilton A. J. S., Gnedin N. Y., 2004, MNRAS, 381, 1
\bibitem[\protect\citeauthoryear{Navarro, Frenk, \& White} {1995}]{nfw95} 
Navarro J.~F., Frenk C.~S., White S.~D.~M., 1995, MNRAS, 275, 720
\bibitem[\protect\citeauthoryear{Navarro, Frenk, \& White} {1996}]{nfw96} 
Navarro J.~F., Frenk C.~S., White S.~D.~M., 1996, ApJ, 462, 563
\bibitem[\protect\citeauthoryear{Navarro, Frenk, \& White} {1997}]{nfw97} 
Navarro J.~F., Frenk C.~S., White S.~D.~M., 1997, ApJ, 490, 493
\bibitem[\protect\citeauthoryear{Oguri et al.} {2008}]{Oguri08}Oguri M. et al., 
2008, AJ, 135, 512
\bibitem[\protect\citeauthoryear{Oguri et al.} {2012}]{Oguri12}
Oguri et al., 2012, AJ, 143, 120
\bibitem[\protect\citeauthoryear{Oh et al.} {2008}]{Oh08}
Oh S.-H., de Blok W. J. G., Walter F., Brinks E., Kennicutt R. C., Jr, 2008, 
AJ, 136, 2761 
\bibitem[\protect\citeauthoryear{Pakmor et al.}{2016}]{Pakmor16}
Pakmor R., Springel V., Bauer A., Mocz P., Munoz D. J., Ohlmann S. T., 
Schaal K., Zhu, C., 2016, MNRAS, 455, 1134
\bibitem[\protect\citeauthoryear{Parry, Eke \& Frenk}{2009}]{PEF09}Parry O. H., 
Eke V. R., Frenk C. S., 2009, MNRAS, 396, 1972
\bibitem[\protect\citeauthoryear{Press \& Schechter} {1974}]{press74} Press W. 
H., Schechter P., 1974, ApJ, 187, 425



\bibitem[\protect\citeauthoryear{Reed et al.} {2007}]{Reed07} Reed D. S., Bower 
R., Frenk C. S., Jenkins A., Theuns T., 2007, MNRAS, 374, 2



\bibitem[\protect\citeauthoryear{Remus et al.}{2013}]{Remus13}
Remus R.-S., Burkert A., Dolag K., Johansson P. H., Naab T., Oser L., Thomas J., 2013, ApJ, 766, 71

\bibitem[\protect\citeauthoryear{Reyes et al.}{2011}]{Reyes11}
Reyes R., Mandelbaum R., Gunn J. E., Pizagno J., Lackner C. N., 2011, MNRAS, 417, 2347

\bibitem[\protect\citeauthoryear{Sarbu, Rusin, \& Ma} {2001}]{Sarbu01}
Sarbu N., Rusin D., Ma C. -P., 2001, ApJ, 561, L147
\bibitem[\protect\citeauthoryear{Schaller et al.}{2015}]{Schaller15}
Schaller M. et al., 2015, MNRAS, 451, 1247
\bibitem[\protect\citeauthoryear{Schaller et al.}{2015a}]{Schaller15a}
Schaller M. et al., 2015a, MNRAS, 452, 343

 \bibitem[\protect\citeauthoryear{Schlegel et al.} {1998}]{Schlegel98} 
Schlegel D. J., Finkbeiner D. P., Davis M., 1998, ApJ, 500, 525
\bibitem[\protect\citeauthoryear{Schneider et al.} {1992}]{schne} Schneider P., 
Ehlers J., Falco E. E., 1992, Gravitational Lenses (Berlin: Springer-Verlag)
\bibitem[\protect\citeauthoryear{Shao et al.}{2012}]{SGTF12}Shao S., Gao L., 
Theuns T., Frenk C. S., 2012, MNRAS, 430, 2346
\bibitem[\protect\citeauthoryear{Shapiro, Iliev \& Raga}{1999}]{SIR99}Shapiro 
P. R., Iliev I. T., Raga A. C., 1999, MNRAS, 307, 203



\bibitem[\protect\citeauthoryear{Sheth \& Tormen}{1999}]{Sheth99}
Sheth R. K., Tormen G., 1999, MNRAS, 308, 119

\bibitem[\protect\citeauthoryear{Springel}{2010a}]{Springel10a}
Springel V., 2010a, ARAA, 48, 391
\bibitem[\protect\citeauthoryear{Springel}{2010b}]{Springel10b}
Springel V., 2010b, MNRAS, 401, 791
\bibitem[\protect\citeauthoryear{Tasitsiomi et al.} {2004}]{Tas04} Tasitsiomi
A., Kravtsov A. V., Wechsler R. H., Primack J. R., 2004, ApJ, 614, 533

\bibitem[\protect\citeauthoryear{Tescari et al.}{2014}]{Tescari14}Tescari E., 
Katsianis A., Wyithe J. S. B., Dolag K., Tornatore L., Barai P., Viel M., 
Borgani S., 2014, MNRAS, 438, 3490
\bibitem[\protect\citeauthoryear{Teyssier}{2002}]{Teyssier02}
Teyssier R., 2002, A\&A, 385, 337
\bibitem[\protect\citeauthoryear{Tinker et al.} {2008}]{Tinker08} Tinker J., 
Kravtsov A. V., Klypin A., Abazajian K., Warren M., Yepes G., Gottlober S., 
Holz D. E., 2008, ApJ, 688, 709



\bibitem[\protect\citeauthoryear{Warren et al.} {2006}]{Warren06} Warren M. S., 
Abazajian K., Holz D. E., Teodoro L., 2006, ApJ, 646, 881


\bibitem[\protect\citeauthoryear{Watson et al.} {2013}]{Watson13} Watson W. 
A., Iliev I. T., D'Aloisio A., Knebe A., Shapiro P. R., Yepes G., MNRAS, 433, 
1230


\bibitem[\protect\citeauthoryear{White \& Rees}{1978}]{WR78}White S. D. M., 
Rees M. J., 1978, MNRAS, 183, 341

\bibitem[\protect\citeauthoryear{Xu et al.}{2016}]{Xu16}
Xu D., Springel V., Sluse D., Schneider P., Sonnenfeld A., Nelson D., 
Vogelsberger M., Hernquist L., 2016, peprint (arXiv: 1610.07605)
 
\bibitem[\protect\citeauthoryear{Yannick et al.}{2017}]{Yannick17}
Yannick M. B. et al., 2017, preprint (arXiv: 1703.10610)


\bibitem[\protect\citeauthoryear{Yang \& Chen}{2009}]{YC09}Yang X.-J., Chen 
D.-M., 2009, MNRAS, 394, 1449 
\bibitem[\protect\citeauthoryear{Yang et al.} {2003}]{Yang03} Yang X., Mo H.
J., van den Bosch F. C., 2003, MNRAS, 339, 1057
\bibitem[\protect\citeauthoryear{Zhang} {2004}]{Zhang04} Zhang T. -J., 2004, 
ApJ, 602, L5

\end{thebibliography}
\end{document}